\documentclass[showpacs,preprint,preprintnumbers,amsmath,amssymb]{revtex4}
 \usepackage{dcolumn}
 \usepackage{bm}

\begin{document}

 \preprint{}

 \title{Topological Properties of Spatial Coherence Function}

 \author{Ji-Rong Ren }
 \author {Tao Zhu }\thanks{Corresponding author. Email : zhut05@lzu.cn }
\author{Yi-Shi Duan}
\affiliation{Institute of Theoretical Physics, Lanzhou University, Lanzhou
730000, P. R. China}

 \date{\today}

 \begin{abstract}
Topology of the spatial coherence function is considered in details. The phase
singularity (coherence vortices) structures of coherence function are
classified by Hopf index and Brouwer degree in topology. The coherence flux
quantization and the linking of the closed coherence vortices are also studied
from the topological properties of the spatial coherence function.
\end{abstract}



 \maketitle

 \section{Introduction}
Coherence function, which was first introduced by Wolf in 1954, is a
very important quantity in describing the cross correlation between
the fluctuating fields at different spacetime points in coherence
optics\cite{1}. Due to its theoretical importance and practical
interest, the concept of optical coherence has been studied
extensively, and considerable progress has been made during the past
five decades. More recently, the existence of coherence vortices or
the phase singularities of a complex coherence function has been
theoretically predicted, and examined by experiments\cite{2,3,4,5}.
The intriguing characteristics of coherence vortex have drawn great
interest. A great deal of works on the coherence vortex have been
done by many physicists\cite{2,3,4,5}.

On the other hand, the existence of the coherence vortices is a topological
phenomena for the complex coherence function, but their topological properties
are not clear. The coherence vortex is a new type of topological objects. In
three dimensional space, these topological objects form string-like vortices,
i.e.,the coherence vortex lines. In a general case, these vortex lines could be
closed, linked together, and even knotted. The coherence vortices carry rich
topological information. It is what inspires us to use topological method to
study the coherence vortices.

In this paper, by making use of the $\phi$-mapping method
\cite{topo,knot,vorticity,opt}, we study the topological current of coherence
vortex and topological knot invariant of the knotted coherence vortex lines in
detail. Meanwhile, we also investigate the coherence flux quantization of the
coherence vortex. This paper is arranged as follows. In Sec.II, we construct a
coherence vorticity field. The topological coherence current of the coherence
vortex lines can be naturally introduced from this coherence vorticity field.
The topological coherence current don't vanish only when the coherence vortex
lines exist. The topological charges of these vortex lines are expressed by the
topological quantum number, the Hopf index and Brouwer degree of the
$\phi$-mapping. In Sec. III, we introduce the coherence flux by analogy to the
magnetic flux, and find that this coherence flux is quantized in topological
level, which is similar to magnetic flux quantization in superconductor and
superfluid systems\cite{vorticity}. Then we research the closed and knotted
coherence vortex lines. An exact expression of the topological knot invariant
is given. Sec. IV is our concluding remarks.

\section{Topological coherence current and coherence vortex}
In the theory of coherence vortex, the coherence current is a new
concept which was introduced in Ref.\cite{5}. For the spatial
coherence function, the coherence current density can be expressed
as
\begin{eqnarray}
\vec{T}(\vec{x}_1,\vec{x}_2)=Im[\Gamma^*(\vec{x}_1,\vec{x}_2) \nabla_1
 \Gamma(\vec{x}_1,\vec{x}_2)] \label{coherence current},
\end{eqnarray}
where $\Gamma(\vec{x}_1,\vec{x}_2)$ is the mutual spatial coherence function,
and $\nabla_1$ is the differential operations to be performed with respect to
the point $\vec{x}_1$. In Eq.(\ref{coherence current}), we see that there are
two spatial variables $\vec{x}_1$ and $\vec{x}_2$. In fact the coherence domain
is just the one of variable spaces for $\vec{x}_1$ or $\vec{x}_2$ , in our
discussions, we will restrict to the case for the variation of location
$\vec{x}_1$ by keeping $\vec{x}_2$ fixed.

The complex mutual coherence function $\Gamma(\vec{x}_1,\vec{x}_2)$ can be
written as
\begin{equation}
\Gamma(\vec{x}_1,\vec{x}_2)=\phi^1+i\phi^2 ,\label{coherence function}
\end{equation}
in which $\phi^a$ $(a=1,2)$ are real functions. Using this $\phi$-field, we can
define an unit vector field: $n^a=\frac{\phi^a}{\|\phi\|}$
$(\|\phi\|^2=\phi^a\phi^a=\|\Gamma\|^2,
   a=1,2)$. By using of this $n$-field, the coherence current density
in Eq.(1) can be expressed as
\begin{equation}
\vec{T}(\vec{x}_1,\vec{x}_2)=\|\Gamma\|^2 \epsilon_{ab}n^a\nabla_1
n^b.\label{coherence current 2}
\end{equation}
For convenience, we will instead $\vec{x}_1$ by variation $\vec{x} $ in our
following discussions.

   By analogy to fluid mechanics system, the coherence velocity field
associated with the $n$-field can be defined as $\vec{u}=\epsilon_{ab}n^a\nabla
n^b$, then the coherence vorticity is given by
$\vec{\omega}=\frac{1}{2}\nabla\times \vec{u}$. This coherence vorticity is a
very essential topological quantity in our discussion, the topological
interpretation of this vorticity is due to the existence of coherence vortices,
and only in this case the vorticity does not vanish\cite{vorticity}. So the
exact expression for $\vec{\omega}$ plays an essential role in topology, in our
following discussions, we will show what the exact expression for
$\vec{\omega}$ is. By using of $n$-field, it can be expressed as
\begin{equation}
\omega^i= \frac{1}{2} \epsilon^{ijk}\epsilon_{ab}\partial_j n^a
\partial_k n^b.\label{topological coherence}
\end{equation}
Obviously, this is a topological current (which we term the topological
coherence current). This current $\omega^i$ play an important role in the
topological properties of the coherence optical systems. Here we will use the
$\phi$-mapping method to research the properties of this current. Using
$\partial_\mu(\phi^a/\|\phi\|)=(\partial_\mu\phi^a)/\|\phi\|+\phi^a\partial_\mu(1/\|\phi\|)$
and the Green function relation in $\phi$ space: $\partial_a
\partial_a \ln \|\phi\| =2\pi
\delta^2(\vec{\phi})~~(\partial_a=\partial/\partial\phi^a)$, we can find a
non-zero expression for $\omega^i$\cite{topo}:
\begin{equation}
\omega^i=\delta^2(\vec{\phi})D^i\left(\frac{\phi}{x}\right),\label{topological
current}
\end{equation}
where the Jacobian vector is defined as
$$\epsilon^{ab}D^i\left(\frac{\phi}{x}\right)
=\epsilon^{ijk}\partial_j\phi^a\partial_k\phi^b. $$

From the expression in Eq. (5), we can come to an important conclusion that the
topological coherence current $\omega^i$ does not vanish only when
$\vec{\phi}=0$. The zeros of $\vec{\phi}$ determine the topological properties
of the topological coherence current $\omega^i$, i.e., the coherence vorticity.
So it is necessary to study the zero points of $\vec{\phi}$ to determine the
nonzero solutions of $\omega^i$. The implicit function theory \cite{imf} shows
that under the regular condition
$$D^i\left(\frac{\phi}{x}\right)\neq 0, $$
the general solutions of
\begin{equation}
\phi^1(x^1, x^2, x^3)=0~,~~~\phi^2(x^1, x^2, x^3)=0
\end{equation}
can be expressed as
\begin{equation}
\vec{x}=\vec{x}_k(s),
\end{equation}
which represent the $N$ isolated singular strings $L_j$ $(j=1, 2, \ldots , N)$
with parameter $s$. These singular string solutions are just the coherence
vortex lines, i.e., the phase singularities of spatial coherence function.
These coherence vortex lines have the coherence function equal to zero, and the
phase of coherence function undefined. In the core of the coherence vortices,
the intensities of the wavefield do not vanish, this is quite different from
the traditional optical vortices in the coherence field.

In $\delta$-function theory\cite{del}, one can prove that in three dimensional
space
\begin{equation}
\delta^2(\vec{\phi})=\sum_{k=1}^{N} \beta_k \int_{L_k}
\frac{\delta^3(\vec{x}-\vec{x}_k(s))}{\mid D(\frac{\phi}{u})\mid_{\Sigma_k}
 }
ds,
\end{equation}
where $D(\phi/u)_{\Sigma_k}=\frac{1}{2}\epsilon^{jk}\epsilon_{mn}(\partial
\phi^m/\partial u^j)(\partial \phi^n/\partial u^k)$, and $\Sigma_k$ is the
$k$th planar element transverse to $L_k$ with local coordinates $(u^1, u^2)$.
The positive integer $\beta_k$ is the Hopf index of $\phi$-mapping, which means
that when $\vec{x}$ covers the neighborhood of the zero point $\vec{x}_k(s)$
once, the vector field $\vec{\phi}$ covers the corresponding region in $\phi$
space $\beta_k$ times. Meanwhile taking notice of the definition of the
Jacobian, one can obtain the direction vector of $L_k$
\begin{equation}
\frac{dx^i}{ds}\Big|_{x_k}=\frac{D^i(\phi/x)}{D(\phi/u)}\Big|_{x_k}.
\end{equation}
Then from Eq. (8) and (9) we obtain the topological inner structure of the
topological coherence current $\omega^i$:
\begin{equation}
\omega^i=\delta^2(\vec{\phi})D^i\left(\frac{\phi}{x}\right)=\sum_{k=1}^{N}W_k
\int_{L_k} \frac{dx^i}{ds} \delta^3(\vec{x}-\vec{x}_k(s))ds,
\end{equation}
in which $W_k=\beta_k \eta_k$ is the winding number of $\vec{\phi}$ around
$L_k$, with $\eta_k=$sgn$ D(\phi/u)_{x_k}=\pm1$ being the Brouwer degree of
$\phi$-mapping. The signs of Brouwer degree are very important, the $\eta_k=+1$
corresponds to the vortex, and $\eta_k=-1$ corresponds to the anti-vortex. The
integer number $W_k$ measures windings of the phase around the phase
singularities of the spatial coherence function, and is called the topological
charge of the coherence vortices. Hence the topological charge of the coherence
vortex line $L_k$ is\cite{topo}
\begin{equation}
Q_k=\int_{\Sigma_k} \omega^i d\sigma_i=W_k.
\end{equation}
By analogy to the optical vortices whose topological charge play the role of an
angular momentum, here the topological charge of coherence vortex also
associated with angular momentum, this is the unique characteristics for the
generic vortices.

In the above discussion, we have known that the coherence vortices are zeros of
the coherence function. The coherence current $\vec{T}$ and the topological
coherence current $\omega^i$ are both expressed by use of the coherence
function. In the theory of coherence function, there is another important
quantity which called spectral degree of coherence and denoted
$\mu(x_1,x_2)$\cite{2,3}. This spectral degree of coherence is a measure of the
spatial coherence of the optical wavefield and takes on values between $0$ and
$1$, zero representing complete incoherence, unity representing complete
coherence. As theoretically predicted, the coherence vortices have the spectral
degree of coherence equal to zero, so the coherence vortices can also derive
from the topological coherence current defined by $\mu(x_1,x_2)$. The
discussions by use of $\mu(x_1,x_2)$ are very similar with the discussions by
coherence function.

 \section{coherence flux and geometry of coherence vortex lines  }
In this section we begin to study the coherence flux and geometry structure of
the coherence vortex lines in three dimensional coherence domain.

From the definition of the coherence vorticity and by analogy to the definition
of magnetic flux, the coherence flux of $\vec{\omega}$ through a surface $S$
can be defined as
\begin{equation}
\Phi=\int_S\vec{\omega}\cdot d\vec{S}.\label{flux}
\end{equation}
It is also called coherence circulation in Ref.\cite{5}. It is known from above
section that $\vec{\omega}$ does not vanish only when the coherence vortices
exist, so the coherence flux associated with the coherence vortex lines through
the surface $S$. According to Eq.(11) and Eq.(\ref{flux}), each coherence
vortex line $L_k$ carries a coherence flux, i.e.,
\begin{equation}
\Phi_k=\int_{\Sigma_k} \omega^i d\sigma_i=W_k=Q_k,
\end{equation}
which will lead to the phenomenon of coherence flux quantization. So the total
flux through a surface $S$ can be expressed as a quantized form
\begin{equation}
\Phi=\sum_{k=1}^{N}W_k=\sum_{k=1}^{N}\beta_k\eta_k=\sum_{k=1}^{N}\Phi_k.
\end{equation}
This is the topological essence of flux quantization. Here, the topological
interpretation is clear, that is the flux can be quantized due to the
topological properties of the coherence function and expressed by the
topological quantum numbers: the Hopf index and Brouwer degree, which are
important topological information carried by the coherence vortices. This is
similar to the topological quantization of the magnetic flux in superconductor
and superfluid systems.

From Eq.(4) we can easily obtain that $\nabla\cdot\vec{\omega}=0$. In the
coherence domain $V$ with a closed surface $S$, by using the Gauss theorem, we
can obtain
\begin{equation}
\Phi=\oint_S\vec{\omega}\cdot d\vec{S}=\int_{V}\nabla\cdot\vec{\omega}dV=0,
\end{equation}
which suggests that the coherence vortex lines are constrained to form closed
loops within the finite spatial volume of the coherence function, or to
terminate at infinity \cite{5}. This closed geometry also appears in fluid
mechanics and classical Electromagnetism, in which the fluid vortices and
magnetic lines are closed.

In fluid mechanics or classical Electromagnetism, the closed vortices or
magnetic lines usually form a knotlike structure\cite{fluid,elec}. Knotted
configurations exist ubiquitously in nature, it also appears as knotted optical
vortices in optical system\cite{optknot1,optknot2,optknot3,optknot4}. For the
closed coherence vortex lines in three dimensional coherence domain, we expect
that they can also form knot structure. Notice that they can be understood as
the linking of two quantized coherence fluxes of the knotted vortex lines. It
is known that for a knot family there are important characteristic numbers to
describe its topology, such as the self-linking and the linking numbers. In
fluid mechanics and classical Electromagnetism, there is an important
topological knot invariant, Helicity, which measures the linking of the closed
lines \cite{fluid,elec}. In the case of coherence vortex lines, the helicity
for the coherence (we term coherence helicity) defined as
\begin{equation}
H=\frac{1}{2\pi}\int \vec{u}\cdot\vec{\omega} dV.\label{16}
\end{equation}
Substituting Eq.(10) into Eq.(\ref{16}), one can obtain
\begin{equation}
H=\frac{1}{2\pi}\sum_{k=1}^{N}\int_{L_k}\vec{u}\cdot d\vec{x}.\label{helicity}
\end{equation}
It can be seen that when these $N$ coherence vortex lines are $N$ closed
curves, i.e., a family of $N$ knots $\gamma_k (k=1,\cdots,N)$,
Eq.(\ref{helicity}) leads to
\begin{equation}
H=\frac{1}{2\pi}\sum_{k=1}^{N}\oint_{\gamma_k}\vec{u}\cdot
d\vec{x}.\label{helicity1}
\end{equation}
This is a very important expression. Consider a transformation of
coherence function $\Gamma$: $\Gamma'=e^{i\theta}\Gamma$, this gives
the U(1) gauge transformation of $\vec{u}$:
$\vec{u}'=\vec{u}+\nabla\theta$, where $\theta\in R$ is a phase
factor denoting the U(1) gauge transformation. It is seen that the
$\nabla\theta$ term in Eq.(\ref{helicity1}) contributes nothing to
the integral $H$ when the coherence vortex lines are closed, hence
the expression (\ref{helicity1}) is invariant under the U(1) gauge
transformation. In the above discussions, we have proved that the
coherence vortices are closed loops in the finite spatial coherence
domain, therefore the integral $H$ is a spontaneous topological
invariant for the coherence vortex lines in the coherence vortex
theory.

It was proved that the integral $H$ is related to the linking and self-linking
number\cite{knot}
\begin{equation}
H=\sum_{k=1}^NW_k^2SL(\gamma_k)+\sum_{k,l=1(k\neq
l)}^NW_kW_lLk(\gamma_k,\gamma_l),
\end{equation}
where $SL(\gamma_k)$ is self-linking number of $\gamma_k$ and
$Lk(\gamma_k,\gamma_l)$ is the Gauss linking number between different knotted
vortex lines $\gamma_k$ and $\gamma_l$. Obviously two coherence fluxes linked
together can not be separated by any continuous deformation of the field
configuration. This provides the topological stability of the knots. Since the
self-linking number and the Gauss linking number are both the invariant
characteristic numbers of the knotted closed curves in topology, $H$ is an
important topological invariant required to describe the knotted coherence
vortex lines in coherence optical systems.
 \section{Conclusion}
First, we introduce the topological coherence current and obtain the inner
structure of the coherence vortices. The coherence vortices have found at the
every zero point of the complex spatial coherence function under the condition
that the Jacobian determinate $D^i(\frac{\phi}{x})\neq 0$, and the topological
coherence current does not vanish only when the coherence vortices exist. One
also shows that the vortex structures are classified by Hopf index and Brouwer
degree in topology. Second, by analogy to the magnetic flux quantization in
superconductor an superfluid systems, we conclude that the coherence flux is
topologically quantized. The coherence flux is quantized due to the topological
properties of the coherence function and can be expressed by the topological
quantum numbers of the coherence vortices. We also conclude that the coherence
vortex lines are constrained to form closed loops within the finite spatial
volume of the coherence function, or to terminate at infinity. This means that
the coherence vortex lines may be knotted and the coherence helicity is a
spontaneous topological invariant for the coherence vortex lines. The exact
expression of the coherence helicity is also given.

 \begin{acknowledgments}
This work was supported by the National Natural Sci- ence Foundation of China.
 \end{acknowledgments}

 \end{document}